\documentclass[%
 aip,
 amsmath,amssymb,
 reprint,%
]{revtex4-1}

\usepackage{graphicx}% Include figure files
\usepackage{dcolumn}% Align table columns on decimal point
\usepackage{bm}% bold math
\usepackage[utf8]{inputenc}
\usepackage[T1]{fontenc}
\usepackage{mathptmx}

\begin{document}

\preprint{AIP/123-QED}

\date{\today}

\title[]{Tuning the electronic band structure of metal surfaces for enhancing high-order harmonic generation}

% Force line breaks with \\

\author{Hicham Agueny}
 \email{hicham.agueny@uib.no}
\affiliation{Department of Physics and Technology, Allegt. 55,
University of Bergen, N-5007 Bergen, Norway %\\This line break forced with \textbackslash\textbackslash
}%

%\date{15 December 2020}

\begin{abstract}
High harmonic generation (HHG) from condensed matter phase holds promise to promote future cutting-edge research in the emerging field of attosecond nanoscopy. The key for the progress of the field relies on the capability of the existing schemes to enhance the harmonic yield and to push the photon energy cutoff to the extreme ultraviolet (XUV, 10-100 eV) regime and beyond towards the spectral "water window" region (282-533 eV). Here, we demonstrate a coherent control scheme of HHG, which we show to give rise to quantum modulations in the XUV region. The control scheme is based on exploring surface states in transition-metal surfaces, and specifically by tuning the electronic structure of the metal surface itself together with the use of optimal chirped pulses. Moreover, we show that the use of such pulses having moderate intensities permits to push the harmonic cutoff further to the spectral water window region, and that the extension is found to be robust against the change of the intrinsic properties of the material. The scenario is numerically implemented using a minimal model by solving the time-dependent Schrodinger equation for the metal surface Cu(111) initially prepared in the surface state. Our findings elucidate the importance of metal surfaces for generating coherent isolated attosecond XUV and soft-x-ray pulses and for designing compact solid-state HHG devices.  
\end{abstract}

\maketitle

\section{INTRODUCTION}
Coherent strong-light matter interaction has revolutionized our understanding of the properties of matter at the level of electrons~\cite{Zewail1988,Zewail1990,Shapiro2000,Corkum2007,Zewail2010,Kockum2019}. It provides powerful tools for coherent control of quantum systems by exploiting light-induced non-linear processes. High-harmonic generation (HHG) is one such process. Due to its coherent aspect, it is routinely exploited in atomic and molecular gases to generate coherent extreme-ultraviolet (XUV) radiation and attosecond pulses~\cite{Corkum2007,Krausz2009,Li2018} as well as to reveal the characteristic spectral and spatial of the electron motion~\cite{Itatani2004,McFarland2008,Uzan2020}. The extension of the process to the condensed matter phase offers wider applicability. In particular, it promises to develop stable and highly efficient XUV and attosecond light sources in compact solid-state devices. This potential is motivated by the recent achievement of HHG from solid-state materials using laser pulses with a peak intensity of the order of one terawatt per cm$^2$. Such an intensity is too weak to generate high-order harmonics in the gas phase. In addition, solid-state HHG promises to promote future cutting-edge research in attosecond nanoscopy~\cite{Ghimire2011,Vampa2015a,Ciappina2017,Vampa2017,Lakhotia2020}. For instance, it has been reported the possibility of accessing the electronic band structure of materials~\cite{Hohenleutner2015,Vampa2015b}, as well as their optical probing as it allows to access the crystal symmetry, interatomic bonding~\cite{Ghimire2011,Vampa2015b,You2016} and phase diagram of the system~\cite{Silva2018,Silva2019}.

The HHG process in atomic and molecular gases is well established and understood on the basis of the semiclassical three-step model~\cite{Corkum1993,Lewenstein1994}: the ionized electrons via tunnelling by means of a strong laser field, which after getting accelerated and thus acquiring extra energy, are driven back to recombine with the parent ion. Due to the presence of the oscillating field the gained energy is converted into high-frequency radiation. The scenario occurs and repeats itself every half an optical cycle of the laser pulse, thus leading to the generation of coherent attosecond pulses. The coherent aspect of the generated pulses is determined by whether the consecutive harmonics are phase-locked and synchronized. In the HHG spectrum, the process manifests by the appearance of harmonic components located at the multiples of the fundamental laser frequency depending on the symmetry of the involved interactions and is characterized by the harmonic cutoff. 

In solid state systems, the process is however more complicated and the intuitive atomic-based models are often incomplete and do not hold for describing properly the solid-state HHG. The complexity is due to the involved interband and intraband mechanisms~\cite{Zaks2012,Vampa2015a,Peter2017,Wang2018,Yue2020}, in which their role for generating high-order harmonics is still under debate~\cite{Yoshikawa2019}.

In general, pushing the high-energy cutoff region to higher photon energies covering the XUV (10-100 eV) and water window (282-533 eV) spectral regimes and enhancing the corresponding yields are one of the challenges of the HHG process~\cite{Teichmann2016,Lara2016,Li2020}. Generating coherent radiation in these spectral regimes, which consists of isolated attosecond pulses (IAPs)~\cite{Hentschel2001,Krebs2013}, is motivated by their potential use for time-resolved applications in chemical and material sciences~\cite{Teichmann2016,Li2020,Schotz2020}. In particular, these IAPs have the potential to localize the coherent motion of a wavepacket at a specific-site in molecules and solid-state systems, thus offering the opportunity for resolving structural changes occurring during photo-chemical and biological processes as well as phase-transition in insulator-to-metal materials with unprecedented temporal resolution.

The last decade has marked the first experimental observations of solid-state HHG~\cite{Ghimire2011,Zaks2012,Vampa2015a,Vampa2015b,Hohenleutner2015,You2016}, which in turn have sparked off extensive theoretical works (e.g.~\cite{Vampa2015c,Luu2016,Jia2018,Tomohiro2019,Yu2019,Orlando2020,Yue2020}) to further understand the origin of the process and to investigate the possibility of enhancing its efficiency. These theoretical considerations were mostly focused on HHG from bulk semiconductors, in which the electron wavepacket is initially delocalised over bulk states. In these works, the generated harmonics were limited to the lower-order region and up to the 25th order, and the used laser intensities, although operate in the non-perturbative regime, cannot be too high to prevent the material from damage. Alternative studies address the HHG from metal surfaces Cu(111) and precisely from the surface state in which the electron wavepacket is initially localized and the control scheme explores the effect of growing ultrathin layers of the insulator NaCl on single-crystal metal surfaces~\cite{Aguirre2016}. 

In this work, we pursue the aim of establishing coherent control schemes with the capability of improving the efficiency of solid-state HHG and towards the generation of coherent XUV and soft-x-ray pulses. At this end, we systematically explore the sensitivity of the HHG process from the transition-metal surface Cu(111) and specifically from surface states to the spectral properties of chirped laser pulses and to that of the electronic structure of the metal surface itself. The particularity of surface states relies on the characteristic of being located at the outermost atomic surface layers, rendering them easier to be manipulated~\cite{Davision1996}. This in turn affects the physical and chemical properties at the interfaces, and that makes the interaction with transition metals (see e.g.~\cite{Stavros1997,Kammler1999,Kumudu2013,Jialu2019}) of particular importance for studying chemical processes~\cite{Stavros1997,Mudiyanselage2013,Behrens2012}. 

We thus explore HHG from surface states for examining the possibility of extending the harmonic cutoff to the XUV and soft-x-ray regions and simultaneously enhancing the corresponding harmonic yields. This is shown to be achieved on the basis of numerical simulations of the time-dependent Schrodinger equations (TDSE) using a minimal model. In particular, we show that tuning the chirp parameters of the pulse allows one to significantly extend the harmonic cutoff to access the spectral water window region, i.e. from the 15th harmonic in the free-chirp case to the 550th harmonic in the case of an optimized chirped pulse having moderate intensities. Further analysis reveals an enhancement of the harmonic yield accompanied by spectral modulations in the XUV photon energy region (i.e. in the range of 69th and 79th harmonics), unlike previous studies, in which the modulations were observed at lower-order harmonics (i.e. 7th and 9th harmonics)~\cite{Kim2019} generated from bulk crystals using a free-chirp pulse. We further show that the observed modulations originate from quantum-path interference~\cite{Zair2008,Auguste2009} of electrons characterized by short and long travel-time, and which leads to the generation of photons having the same final energy. We validate these observations using a simple physical model that captures the quantum aspect of the modulations. These modulations, in turn, are found to exhibit a strong sensitivity to the change of the number of Cu-monolayers, while beyond the XUV-region the HHG spectra remain unchanged. We exploit this sensitivity for coherent control of the HHG process by selectively manipulating the bulk states combined with the use of optimal chirped pulses. Our findings thus indicate the importance of exploring metal surfaces for the generation and characterization of IAPs in the XUV and soft-x-ray regimes, and thus bridging the gap between attosecond and surface sciences. 

This paper is organized as follows. In Sec. \ref{theory} we present our theoretical model based on a one-dimensional (1D)-TDSE, including a short description of our numerical methods for solving the TDSE and for calculating the HHG spectrum. We stress that extensive theoretical works have been carried out using a 1D-model, and which was shown to capture the basic physics involved in an experiment (e.g. Refs.~\cite{Kazansky2009}). In Sec. \ref{results} we present our results of HHG spectra produced by chirped pulses, and explore the role of bulk states to coherently control the HHG process from the surface state of Cu(111) and discuss the physics behind the emerged effects.  We also predict the generation of IAPs in the XUV and soft-x-ray photon energy regions, and which can be characterized via the chirp parameter of the pulse and the metal surface itself. The findings are supported by an analysis of the calculated population of bulk, image and surface states as well as the use of a simple physical model. Finally, in Sec. \ref{conclusions} we summarize our results of HHG. Atomic units (a.u.) are used throughout this paper unless otherwise specified.

\section{THEORY AND COMPUTATIONAL DETAILS}\label{theory}

The electron dynamics induced inside metal surfaces is modelled using a 1D-model with the use of a one-electron pseudo potential as in previous works~\cite{Kazansky2009,Aguirre2016,Catoire2015}. In a 3D-model, the metal surface electron feels the potential in the direction normal to the surface, while it is considered to move freely with the electron momentum $k_{\parallel}$ in the parallel direction. This justifies our restriction to the one-direction motion of the electron, which is here triggered by a linearly polarized laser pulse directed to the (111) direction of the metal surface. On the basis of this 1D-model, the TDSE governing the electron dynamics induced by coherent light pulses is written as 
\begin{equation}\label{tdse}
\Big[-\frac{\nabla_z^{2}}{2} +V_{ion}(z) + H_I(t) - i\frac{\partial}{\partial t}\Big]|\psi(t) \rangle=0,
\end{equation}
where $V_{ion}(z)$ is an effective potential interaction. Here, we use Chulkov potentials to model the electronic structures of the metal surface Cu(111). The potentials are parameterized and have the analytical form~\cite{Chulkov1999}
\begin{subequations}\label{Vion}
\begin{eqnarray}
V_1(z)&=&A_{10} + A_1\cos(\frac{2\pi}{a_s}z),\;\; z<0 \label{Vion1}
\\
V_2(z)&=&-A_{20} + A_2\cos(\gamma z), \;\; 0<z<z_1 \label{Vion2}
\\
V_3(z)&=&A_3\exp[-\alpha(z- z_1)], \;\; z_1<z<z_{im} \label{Vion3}
\\
V_4(z)&=&\frac{\exp[-\lambda(z- z_{im})]-1}{4(z-z_{im})}, \;\; z_{im}<z, \label{Vion4}
\end{eqnarray}
\end{subequations}
where $a_s$ is the bulk interlayer spacing and $z_{im}$ is the position of the image plane. The parameters $A_{10}$, $A_1$, $A_2$ and $\beta$ are independent parameters and are given in the table~\ref{table1} for the metal surface Cu(111). They are adjustable and are obtained using a fitting procedure involving results from the density function calculations and experiments~\cite{Chulkov1999}. While the parameters $A_{20}$, $A_3$, $\alpha$, $z_1$, $\lambda$ and $z_{im}$ are determined from the continuity condition of the potential and its first derivative, and are given according to~\cite{Chulkov1999,So2015}   
\begin{eqnarray}
A_{20}=A_2-A_{10}-A_1; \;\; A_3=-A_{20}-\frac{A_2}{\sqrt{2}}\nonumber\\
z_i=\frac{5\pi}{4\gamma}; \;\; \alpha=\frac{A_2\gamma}{A_3}\sin(z_1\gamma)\nonumber\\
\lambda=2\alpha; \;\; z_{im}=z_1-\frac{1}{\alpha}ln(-\frac{\alpha}{2A_3}).
\end{eqnarray}

\begin{table}
\caption{\label{table1}Adjustable parameters used in the analytical pseudo potential in Eq. (\ref{Vion}) and which is shown in Fig. \ref{fig1}(a). The values are taken from~\cite{Chulkov1999}. All parameters are given in atomic units.}
\begin{ruledtabular}\label{table1}
\begin{tabular}{cccccddd}
$a_s$ & $A_{10}$ & $A_1$ & $A_2$ & $\gamma$ \\
\hline
 3.94 & -0.4371 & 0.1889 & 0.1590 & 2.9416\\
\end{tabular}
\end{ruledtabular}
\end{table}
Note that the pseudo potential in Eq. (\ref{Vion}) is multiplied by a damping function of the form $[1 + \tanh(z)]/2$, similar to that used in~\cite{Aguirre2016}, to ensure the continuity of the potential at the boundaries. The time-dependent interaction $H_I(t)$ in Eq. (\ref{tdse}) is treated in the length gauge and is described within the dipole approximation according to
\begin{equation}\label{tdi}
H_I(t) = -z S(z) F(t) =  S(z)E_0 f(t) \cos(\omega_0 t + \phi(t)),
\end{equation}
where the function $S(z)$ accounts for screening of the electric field inside the Cu(111) surface (see, e.g., Refs.~\cite{Kazansky2009,Aguirre2016}) and takes the form 
\begin{equation}\label{screen}
S(z) =  0.5\{1+\tanh[6z+3\xi)/\xi]\}.
\end{equation}
As in~\cite{Kazansky2009}, the screening length is chosen to be $\xi=$4 and the change of its value does not affect dramatically the results presented in this work. We have performed calculations for $\xi=$10 and 20 and it is found that the basic physics discussed in the context of HHG remains valid. The function $f(t)$ in Eq. (\ref{tdi}) is the pulse envelope which is chosen to be of a cosine square form, $\omega_0$ is the central angular frequency, and $E_0$ the peak amplitude of the laser pulse and is related to the peak intensity via the relation $I_0=E_0^{2}$. The function $\phi(t)$ in Eq. (\ref{tdi}), which is a time-dependent carrier-envelope phase (CEP), determines the form of the chirped pulse. In this work, we use a chirped pulse similar to the one in Ref. \cite{Carrera2007}, whose phase $\phi(t)$ has the following time-varying profile
\begin{equation}\label{chirp}
 \phi(t) =   -\beta \tanh (\frac{t-t_0}{\tau}),
\end{equation}
and the corresponding instantaneous frequency is obtained according to
\begin{equation}\label{wt}
 \omega(t) =  \omega_0 - \frac{d\phi(t)}{dt}=  \omega_0 -\frac{\beta}{\tau} [\cosh^2(\frac{t-t_0}{\tau})]^{-1},
\end{equation}
where the chirp parameters $\beta$ (given in rad) and $\tau$ can be adjusted to control, respectively, the frequency sweeping range and the steepness of the chirped pulse centred at $t_0$. In Eqs. (\ref{chirp}) and (\ref{wt})  the optical phase $\phi(t)$ becomes zero in the case of $\beta=$ 0, and thus we have $\omega(t)=\omega_0$; this defines the chirp-free case.

As in our previous work~\cite{Malek2020}, we calculate the HHG spectrum $H(\omega)$ by carrying out the Fourier transform of the expectation value of the dipole acceleration along the $z$-axis
\begin{equation}\label{H}
H(\omega) = |D_z(\omega)|^2,
\end{equation} 
where $D_z(\omega)$ is defined by
\begin{equation}\label{Dzw}
D_z(\omega) = \frac{1}{\sqrt{2\pi}} \int_{t_i}^{t_f} <D_z(t)> \mathrm{e}^{-i\omega t} dt,
\end{equation} 
and the time-dependent expectation value of the dipole acceleration $<D_z(t)>$ is written as
\begin{equation}\label{Dzt}
D_z(t) = \langle \psi(t)| \frac{\partial V(r)}{\partial z} + F(t) |\psi(t)\rangle.
\end{equation}  

In Eq. (\ref{Dzw}) $t_i$ and $t_f$ define, respectively, the time at which the pulse is switched on and off. Note that the dipole acceleration $<D_z(t)>$ in Eq. (\ref{Dzt}) is convoluted by a window function of a Gaussian form $\exp{[-(t-t_0)^2/(2\sigma^2)]}$ centred at $t_0$ and having the width of $\sigma$=5.77/$\omega_{0}$. 

In our numerical simulations the initial states are obtained by diagonalizing the matrix representation of the Hamiltonian in Eq. (1) in the absence of the laser pulse, and which is constructed using of a sinus-DVR basis~\cite{Lill1982}. The time evolution of the electronic wave function $\psi(t)$, which satisfies the TDSE [cf. Eq. (\ref{tdse})], is solved numerically using a split-operator method combined with a fast Fourier transform (FFT) algorithm. The calculations are carried out in a grid of size $L_z$ = 8190 with the spacing grid $dz$ = 0.25 a.u., i.e. $n_z$=32768 grid points. The time step used in the simulation is $\delta t$=0.02 a.u.. The convergence is checked by performing additional calculations with twice the size of the box and a smaller time step. An absorbing boundary is employed to avoid artificial reflections, but without perturbing the inner part of the wave function. The boundary is chosen to span 10\% of the grid size in the $z$-direction.

%------------Begin FIGURE 1
%\begin{SCfigure}[ht]
\begin{figure}[ht]
\centering
\includegraphics[width=8.5cm,height=6.5cm]{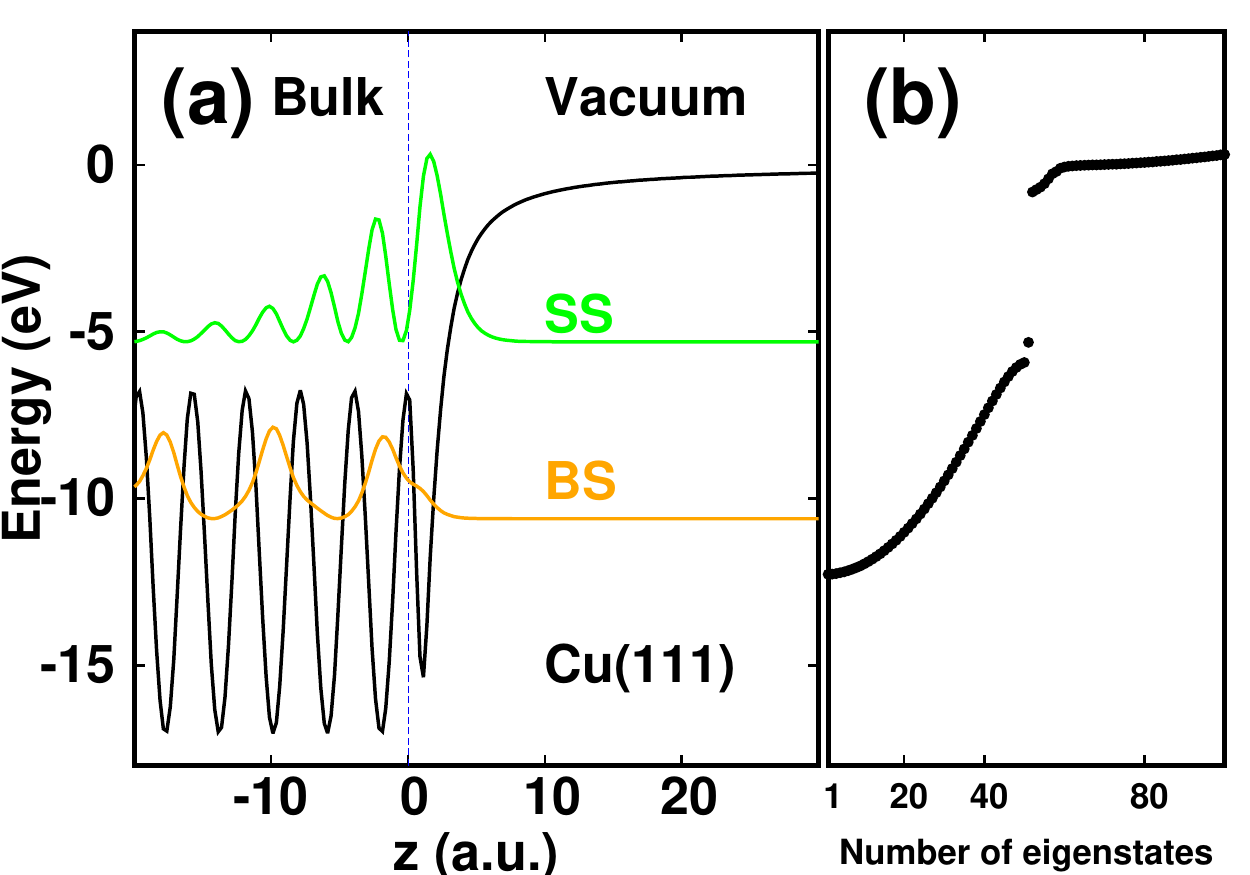}
\caption{\label{fig1}(a) Effective one-dimensional potential (black curve) and the electron densities of the surface state labeled (SS) having the energy -5.3 eV and a bulk state (BS) of energy -10.6 eV for Cu(111). The electron densities of SS and BS are multiplied by 30 and 200, respectively. (b) Eigenenergy at $k_{\parallel}=0$ of the binding electronic states of the model potential in (a). Here $k_{\parallel}$ is the crystal momentum of the electron.}
\end{figure}
%\end{SCfigure}
%------------End FIGURE 1

\section{RESULTS AND DISCUSSION}\label{results}

%------------Begin FIGURE 2
%\begin{SCfigure}[ht]
\begin{figure*}[ht]
\centering
\includegraphics[width=7cm,height=6cm]{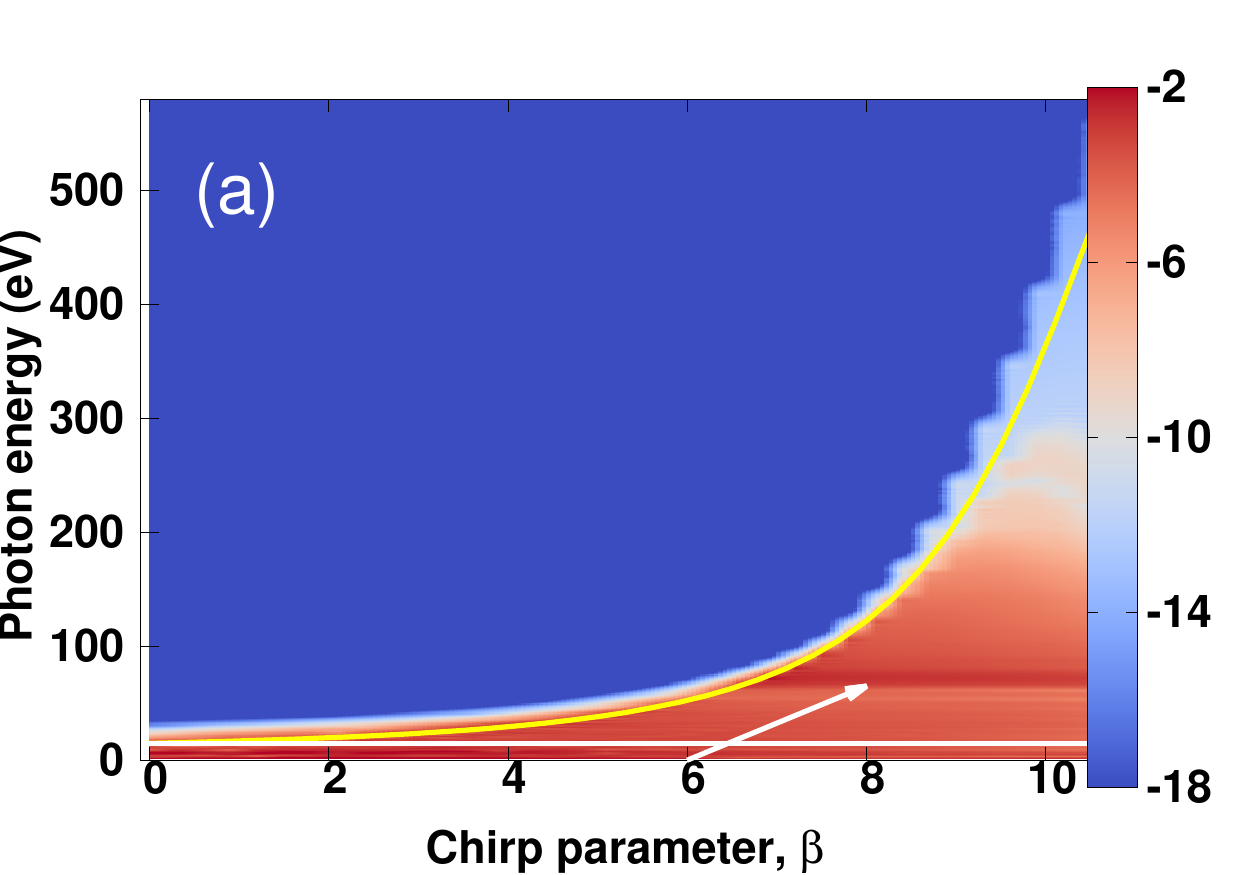}
\includegraphics[width=7cm,height=6cm]{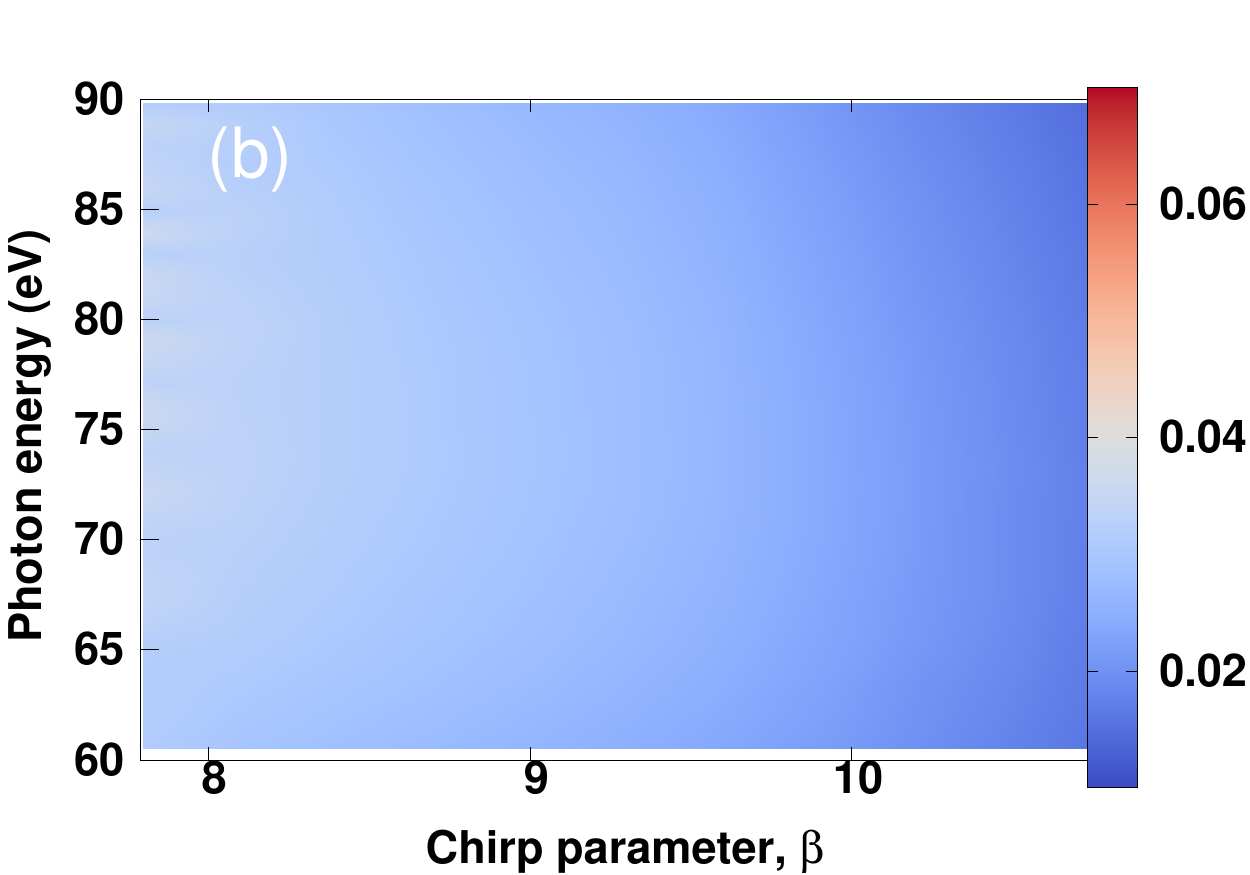}
\includegraphics[width=7cm,height=6cm]{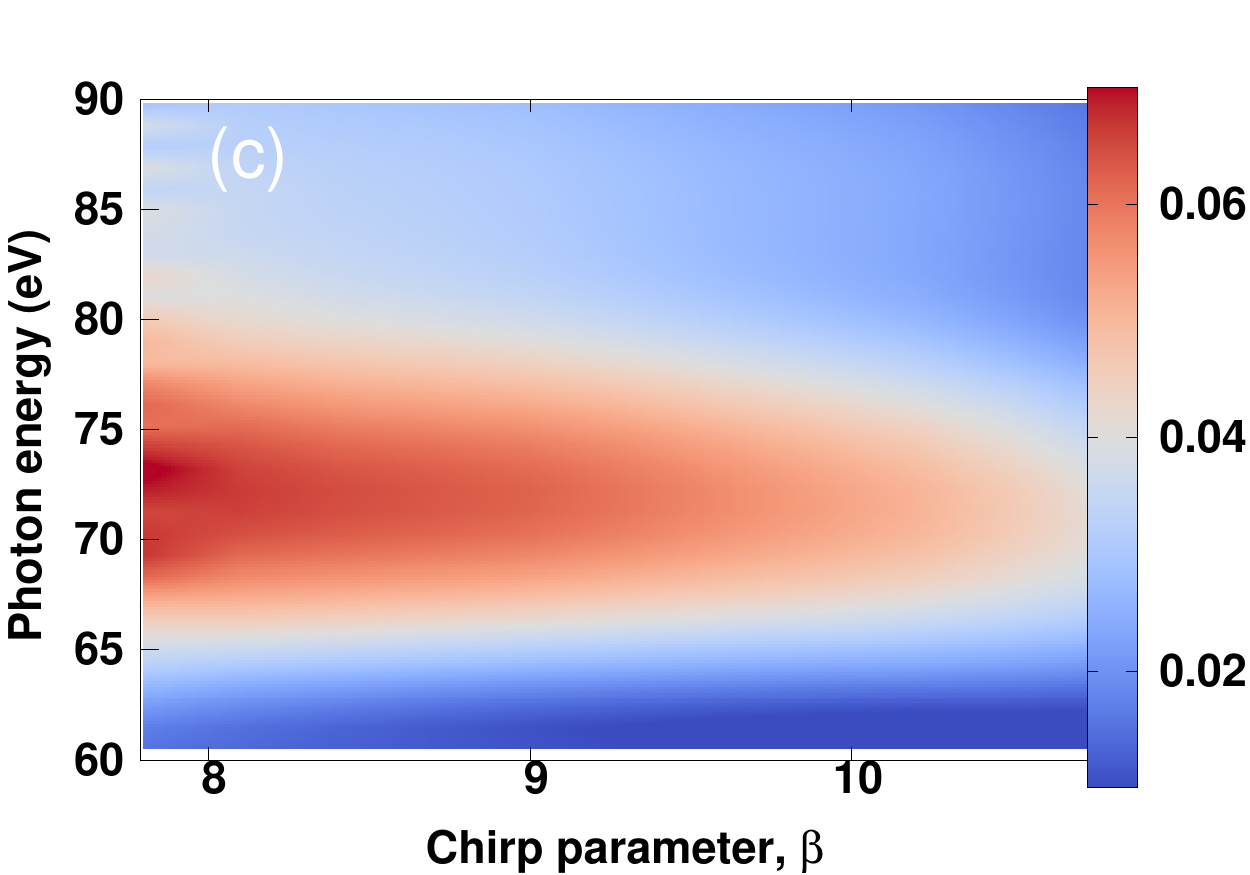}
\includegraphics[width=7cm,height=6cm]{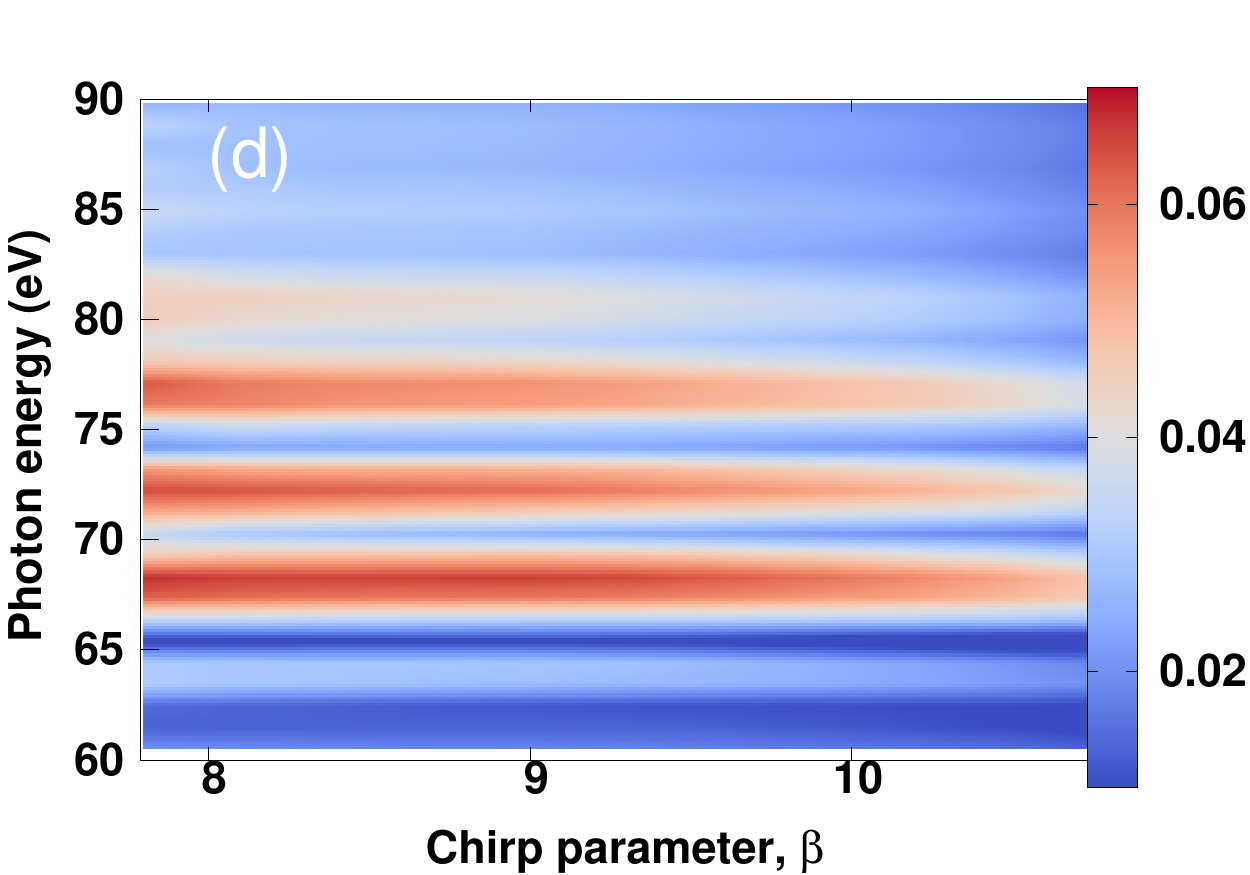}
\includegraphics[width=7cm,height=6cm]{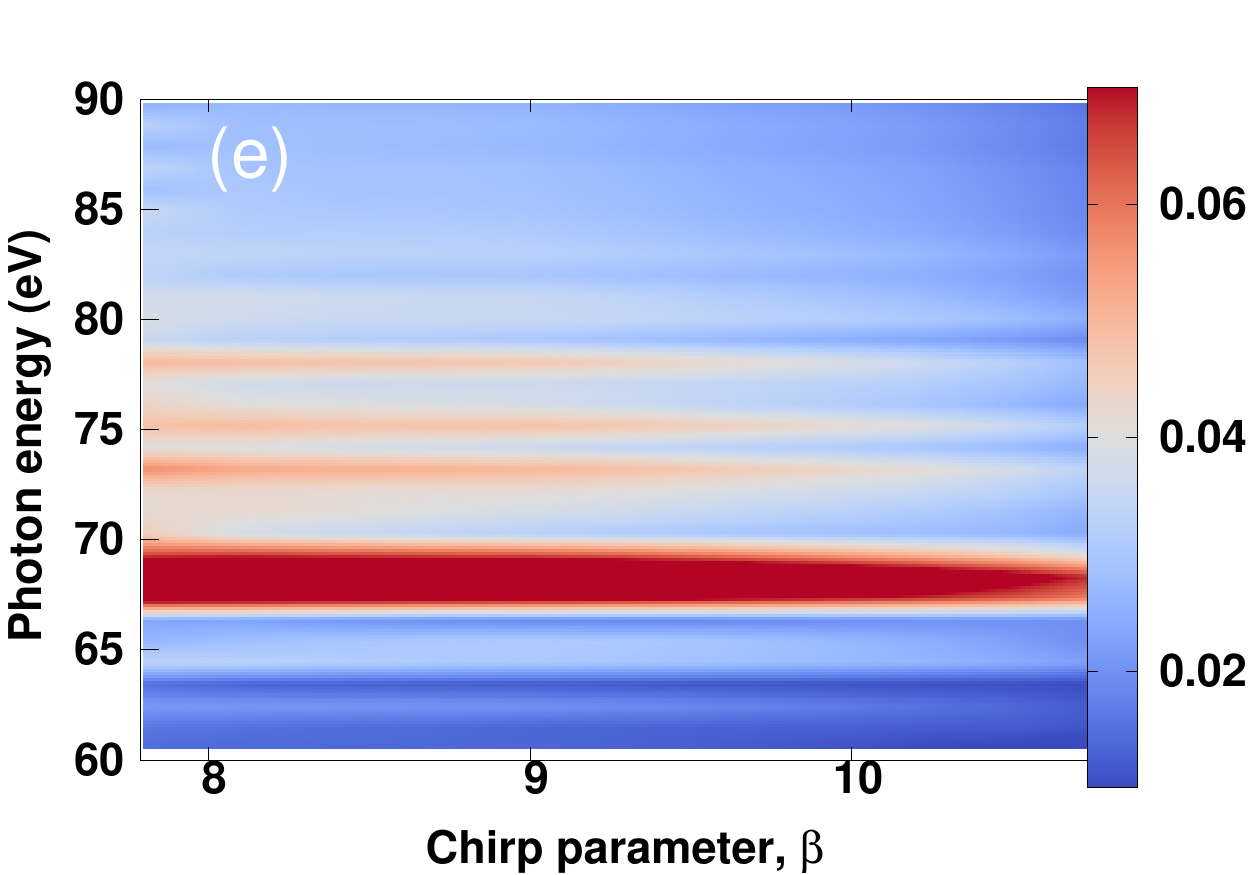}
\includegraphics[width=7cm,height=5.6cm]{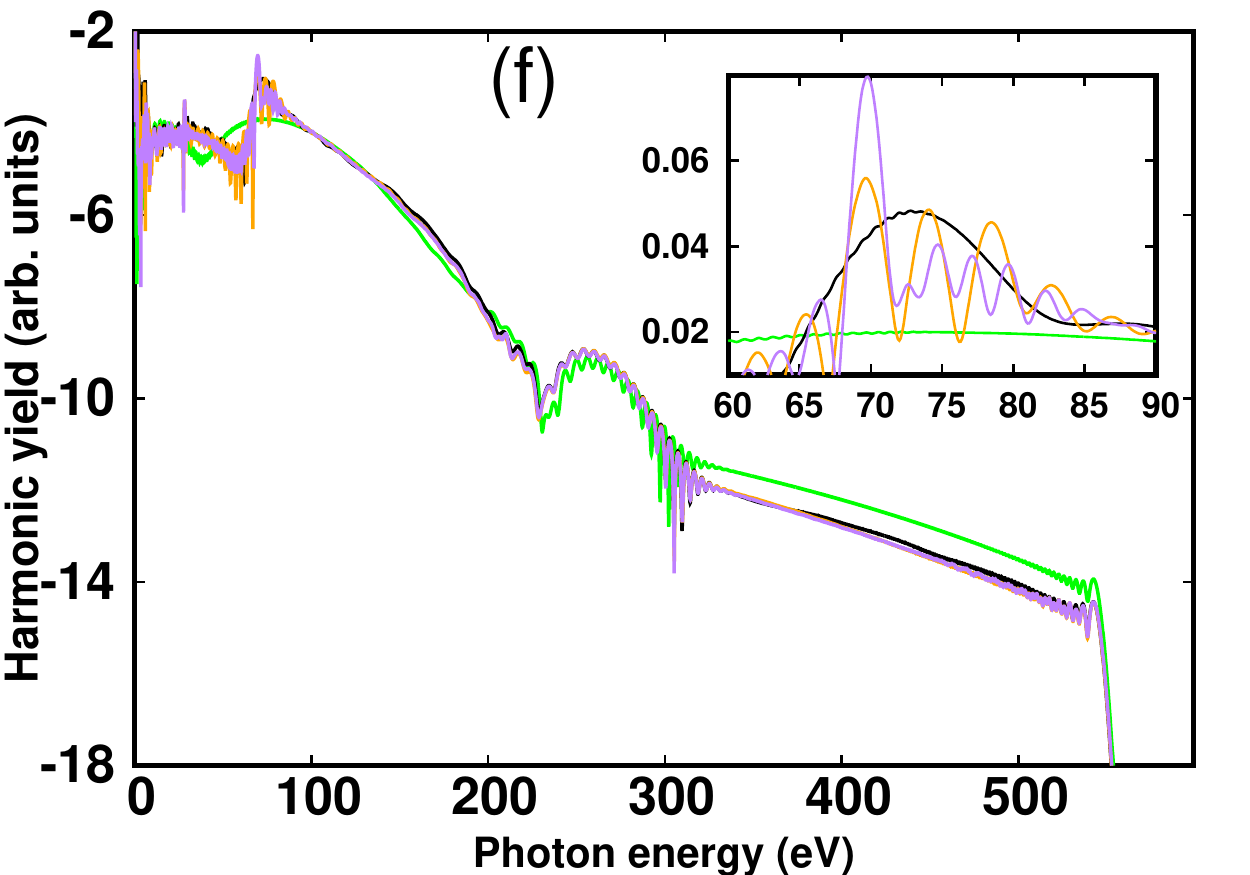}
\caption{\label{fig2} (a) HHG spectra calculated from the Cu(111) surface state with 10 MLs as a function of the chirp parameter $\beta$ in the range 0-10.8 rad. The yellow curve in (a) displays the cutoff energy deduced from the Eq. (\ref{Cutoff}) and the horizontal white line corresponds to the free-chirp case. (b)-(e) The same as (a) but the spectra are displayed in the XUV region as indicated by the arrow in (a) and for $\beta$ in the range 7.7-10.8 rad and at various number of MLs: (b) 1 ML, (c) 10 MLs, (d), 30 MLs and (e) 50 MLs. (f) HHG spectra for $\beta=$10.4 rad at different MLs: 1 ML (green curve), 10 MLs (orange curve), 30 MLs (purple curve) and 50 MLs (black curve). Inset: a zoom of the spectra in (f) in the XUV region. The parameters of the chirped pulse are: $\lambda_{NIR}=$ 1.27 $\mu m$, $T_c=$ 10 cycles, $\tau=$ 300 a.u. and $I_{NIR}=$ 2$\times$10$^{13}$ W/cm$^2$.}
\end{figure*}
%\end{SCfigure}
%------------End FIGURE 2

The control scheme implemented in this work is based on tuning the electronic structure of the metal surface itself combined with the search of optimal parameters of the chirped pulse. The main goal is to investigate the possibility of extending the harmonic cutoff with simultaneous enhancement of the harmonic yield. Here we use the metal surface Cu(111), which is considered to be initially prepared in the surface state having the energy -5.33 eV. The energy is located in the forbidden energy gap and bellow the Fermi level (E$_f$=-4.5 eV), as depicted in Fig. \ref{fig1}(b). The electron density of the corresponding state is shown in Fig. \ref{fig1}(a) (green curve) and exhibits a localized character close to the surface atomic layer and decays into both the vacuum and the bulk~\cite{Davision1996}, unlike bulk states which have a Bloch character and the corresponding electron density is delocalized over the bulk metal and decays into the vacuum, as depicted in Fig. \ref{fig1}(a) (orange curve). We thus exploit this characteristic feature of the surface state for coherent control of the HHG process.

In Fig. \ref{fig2}, we show HHG spectra generated by chirped near-infrared pulses. The pulse is characterized by a chirp parameter $\beta$ and has 1.27 $\mu m$ central wavelength, 43 fs pulse duration and 2$\times$10$^{13}$ W/cm$^2$ as the maximum of the peak intensity. The choice of the pulse intensity is limited by the material damage, as discussed in \cite{Aguirre2016}. Figure \ref{fig2}(a) shows a strong sensitivity of the spectrum to the change of the chirp parameter $\beta$. Here, the parameter $\beta$ covers the region 0-10.8 rad. The choice of these values is such that the instantaneous frequency $\omega(t)$ remains positive. Note that only the positive chirps affect the HHG spectrum. In this region of $\beta$, the energy cutoff extends from $E_{max}=$15 eV for $\beta=$0 (free-chirp case)  to $E_{max}=$540 eV for $\beta=$ 10.4. An extension by almost a factor of 36 is observed. The increase of the energy cutoff $E_{max}$ as a function of increasing the chirp $\beta$ is found to follow the approximative formula
\begin{equation}\label{Cutoff}
E_{max}(\beta)= I_{p} + 3.17 U_p(\beta),
\end{equation}       
as depicted in Fig. \ref{fig2}(a) with white curve. Here $U_p=I/4\omega(\beta)$ is the pondermotive energy of free electrons in an oscillating field and $I_p$ is the ionization potential. The instantaneous frequency $\omega(t)$, which is a time-dependent function [cf. Eq. (\ref{wt})], is found to be determined at times defined as $t = e^{1.45\sqrt{\beta}}$. By introducing this latter formula, the Eq. (\ref{Cutoff}), although  deviates from the well know one in the free-chirp case (see the horizontal white line in Fig. \ref{fig2}(a)), is found to reproduce very well the energy cutoff in the whole range of the chirp parameter. On the other hand, the Eq. (\ref{Cutoff}) captures the basic physics behind the significant extension of the cutoff region. It shows that the free electrons acquire high-kinetic energy from the field, which is inversely proportional to the instantaneous frequency characterized by the chirp parameter. Indeed, increasing the parameter $\beta$ results in decreasing the instantaneous frequency $\omega(t)$, and hence a vast kinetic energy $U_p(\beta)$ up to 170 eV is transferred to the free electrons compared to 3 eV in the free-chirp case. 

%------------Begin FIGURE 3
\begin{figure*}[ht]
\centering
\includegraphics[width=6cm,height=4.6cm]{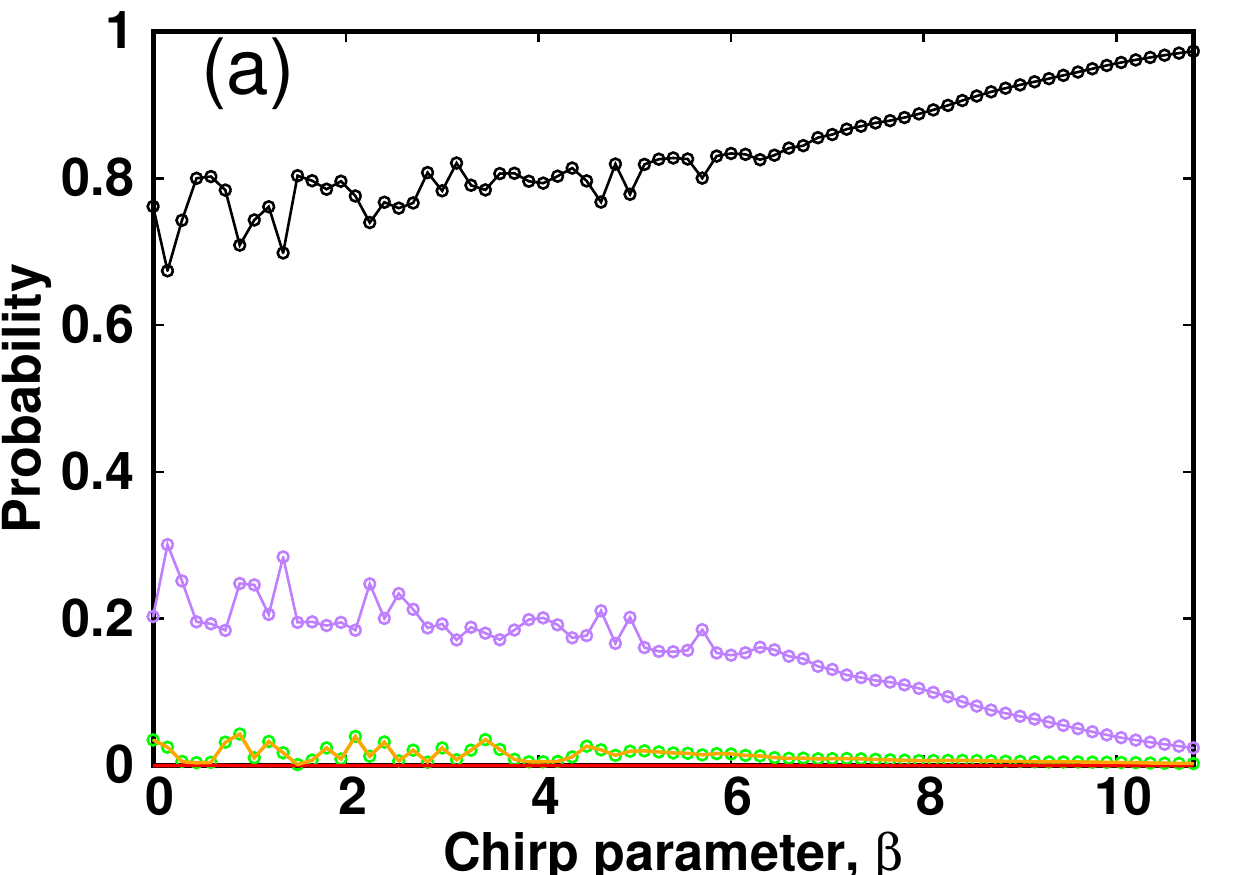}
\includegraphics[width=6cm,height=4.6cm]{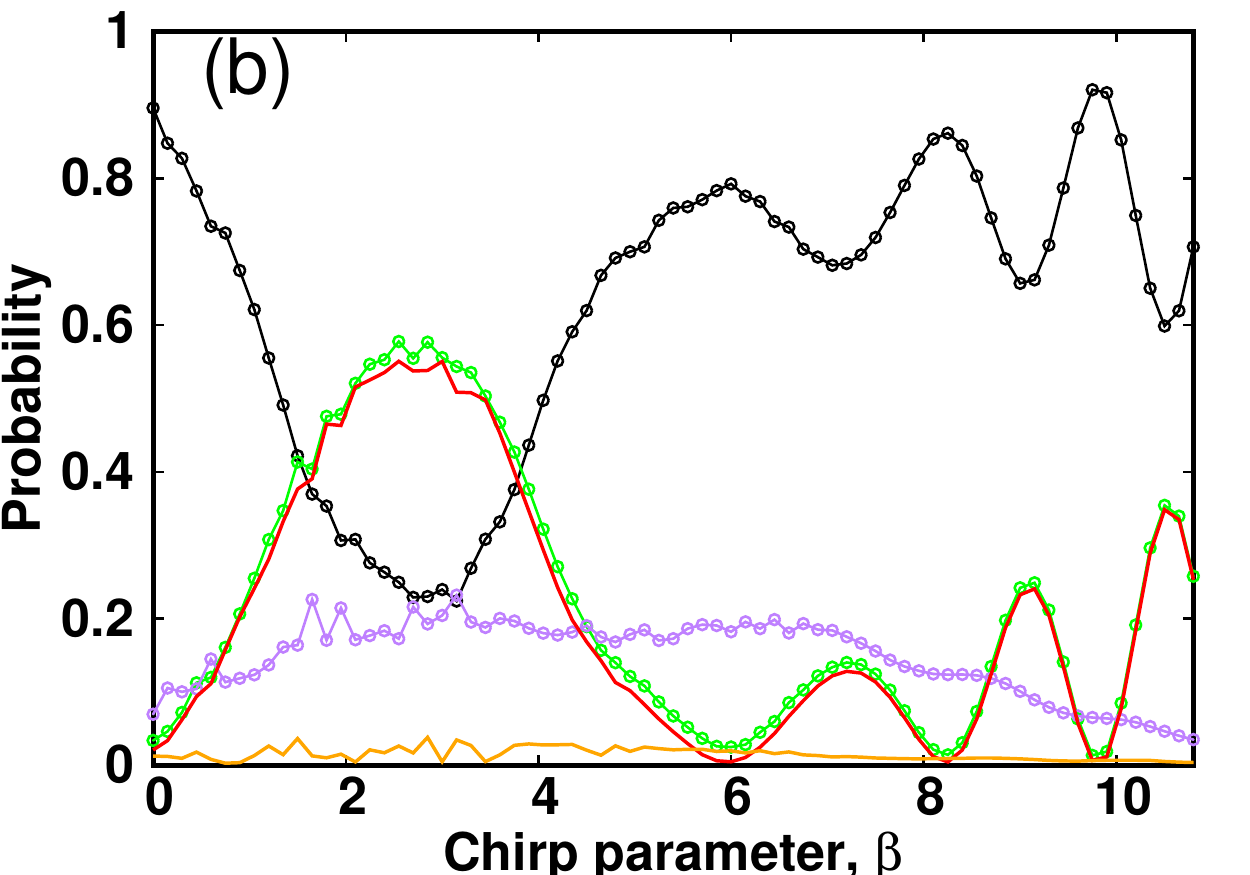}
\includegraphics[width=6cm,height=4.6cm]{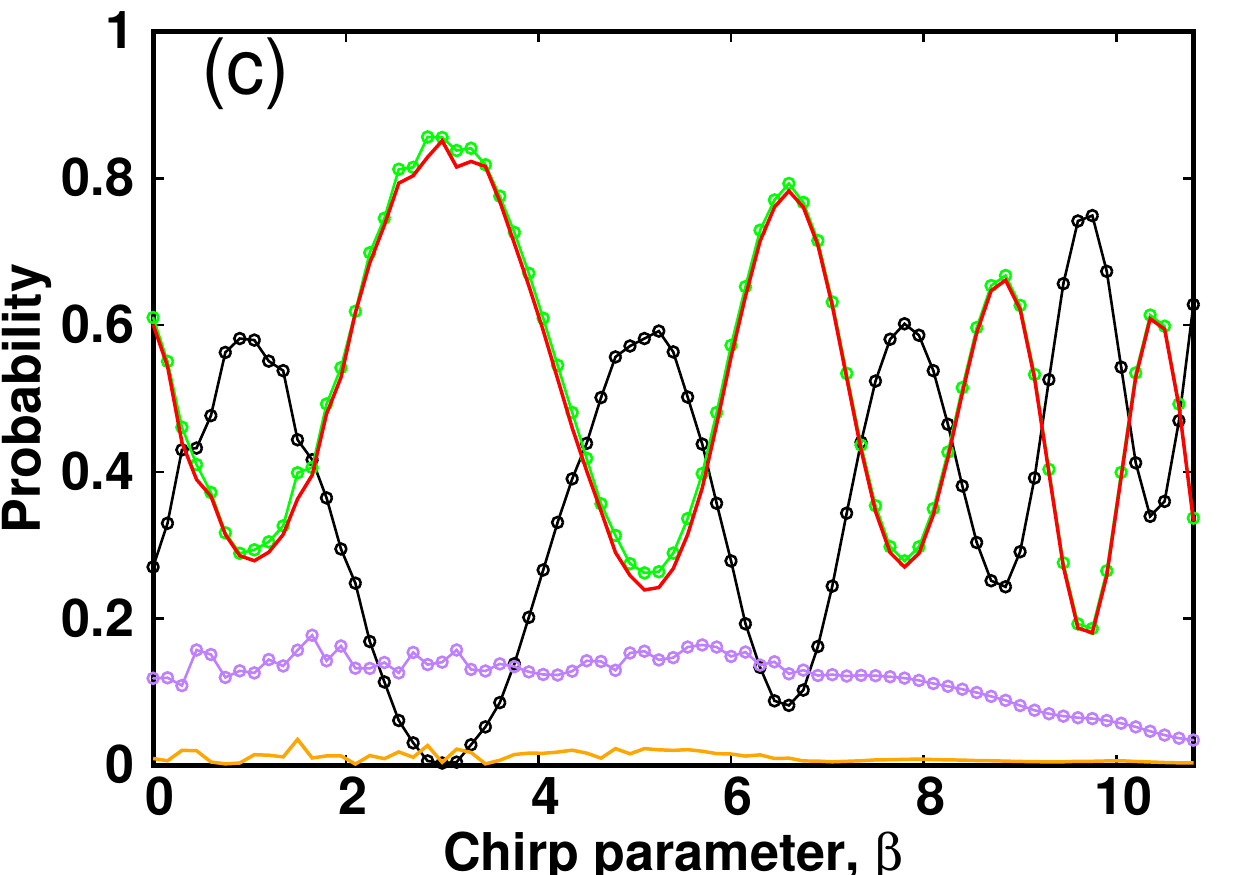}
\includegraphics[width=6cm,height=4.6cm]{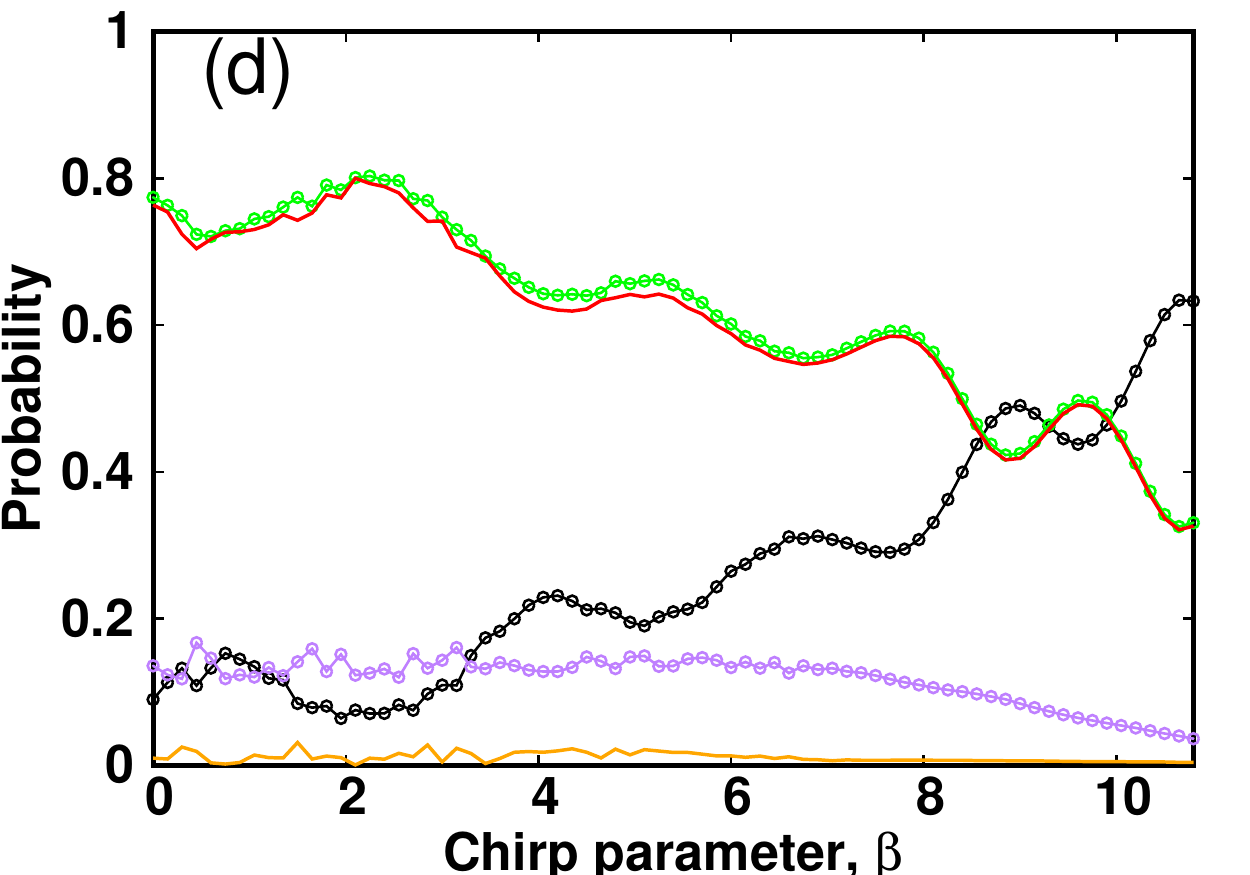}
\caption{\label{fig3}Probability of occupied states as a function of the chirp parameter $\beta$ at various number of MLs: (a) 1 ML, (b) 10 MLs, (c) 30 MLs and (d) 50 MLs. The population is shown for: the initial state (black curve with filled circle), bulk states (red curve) and image states (orange curve). Also are shown the probability of the excitation (green curve with filled circle) and the ionization (purple curve with filled circle). The parameters of the chirped pulse are: $\lambda_{NIR}=$ 1.27 $\mu m$, $T_c=$ 10 cycles, $\tau=$ 300 a.u. and $I_{NIR}=$ 2$\times$10$^{13}$ W/cm$^2$.}
\end{figure*}
%------------End FIGURE 3

A closer inspection of the HHG spectrum shows an enhancement of the harmonic yield in the photon energy region 60-90 eV at the optimal chirp parameter range 7-10.8 rad, as indicated by arrow in Fig. \ref{fig2}(a). The spectra in these ranges are displayed with higher visibility in Figs. \ref{fig2}(b), (c), (d) and (e) at various Cu MLs, respectively, 1, 10, 30 and 50 MLs. Interesting, varying the number of MLs of the metal surface itself results in abrupt change of the spectrum. The change here manifests by spectral modulations, in particular, at 30 and 50 MLs, while these modulations are absent in the case of 10 MLs, in which only an enhancement of the harmonic yield is seen. On the other hand, no changes in the spectrum is observed in the case of 1 ML. This sensitivity of HHG to the change of MLs is mainly seen in the photon energy region 60-90 eV, while beyond this region the curves of the spectrum fall on the top of each other except in the case of 1 ML, in which a slight difference is noticed. On the other hand, the harmonic cutoff remains unchanged as can be seen in Fig. \ref{fig2}(f). In this figure, the spectra are shown for the chip parameter $\beta=10.4$ rad, and a zoom of the spectrum in the XUV photon energy region is displayed as an inset to highlight the emerged modulations.   

These results indicate that by properly chosen the chirp parameter one can produce high-energy photons covering the spectral water range 283-533 eV. On the other hand, tuning the electronic band structure by varying the number of MLs results in an enhancement of the harmonic yield in the XUV photon energy region accompanied by spectral modulations. Such an enhancement is only possible when introducing an optimal chirped pulse, which here plays a role of switching on/off the access to bulk states. As a result, manipulating these states by tuning the MLs of the metal surface itself leads to a dramatic change of the HHG process. Here increasing the number MLs leads to the generation of a large numbers of bulk states, which get involved in the dynamics induced by the chirped pulse. This can be seen in Fig. \ref{fig3}, in which we show the population of the bulk states (red curve), surface state (black curve) and image states (orange curve) as well as the probability of the total excitation (green curve) and the ionization (purple curve) as functions of the chirp parameter $\beta$. These results are presented at different MLs: 1, 10, 30 and 50 MLs. At first glance, the excitation exhibits a strong sensitivity to the change of the number of MLs. Note that in the case of 50 MLs, there are 50 bulk states located below the surface state, and only 10 bulk states in the case of 10 MLs, while no bulk state exists when only 1 ML is considered. Here, the total excitation results from the contribution of bulk states, which is significant except in the case of 1 ML, in which the excitation comes from the contribution of image states. On the contrary, the ionization probability is slightly affected by the change of the number of MLs, which indicate that the implemented control scheme modifies mainly the phase dipole emission. 

\begin{figure*}[ht]
\centering
\includegraphics[width=6.cm,height=4.6cm]{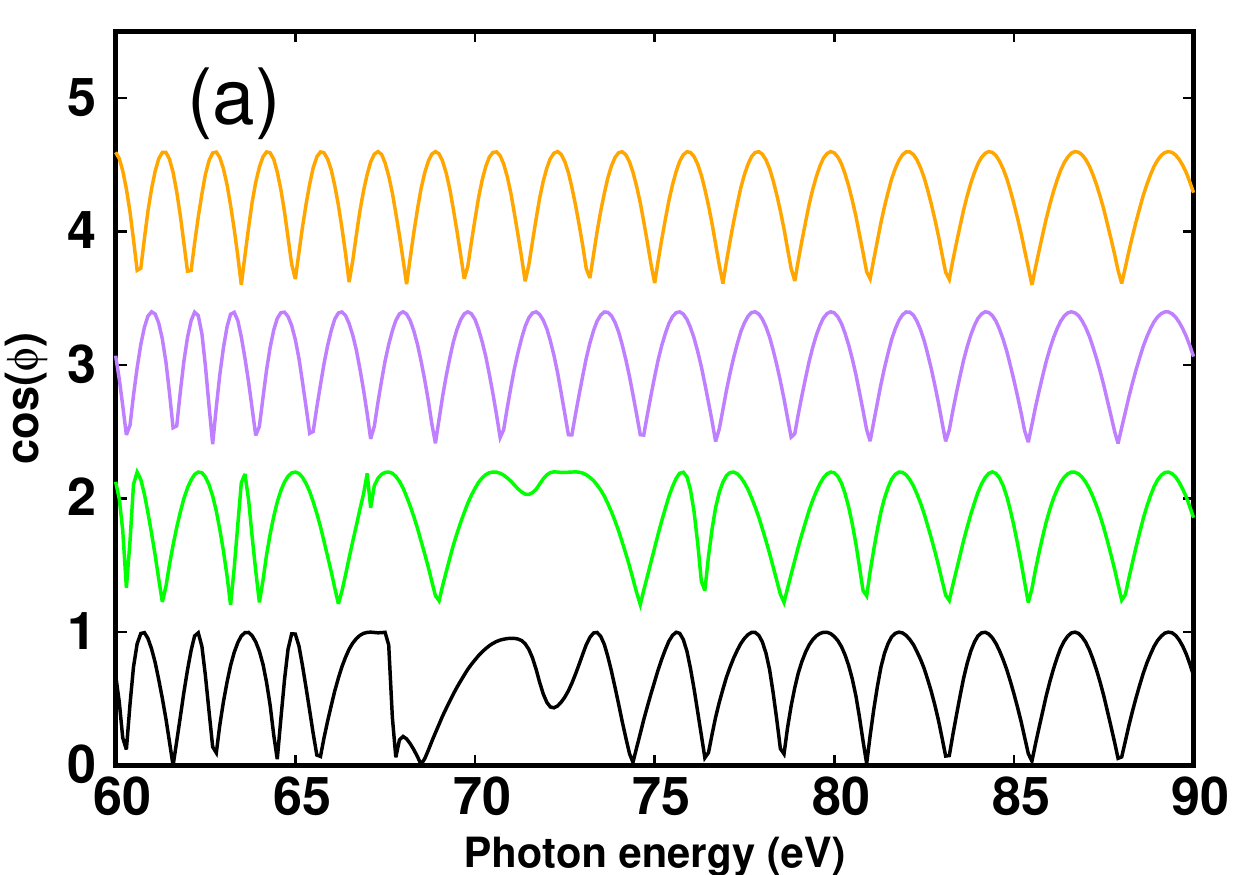}
\includegraphics[width=6.cm,height=4.6cm]{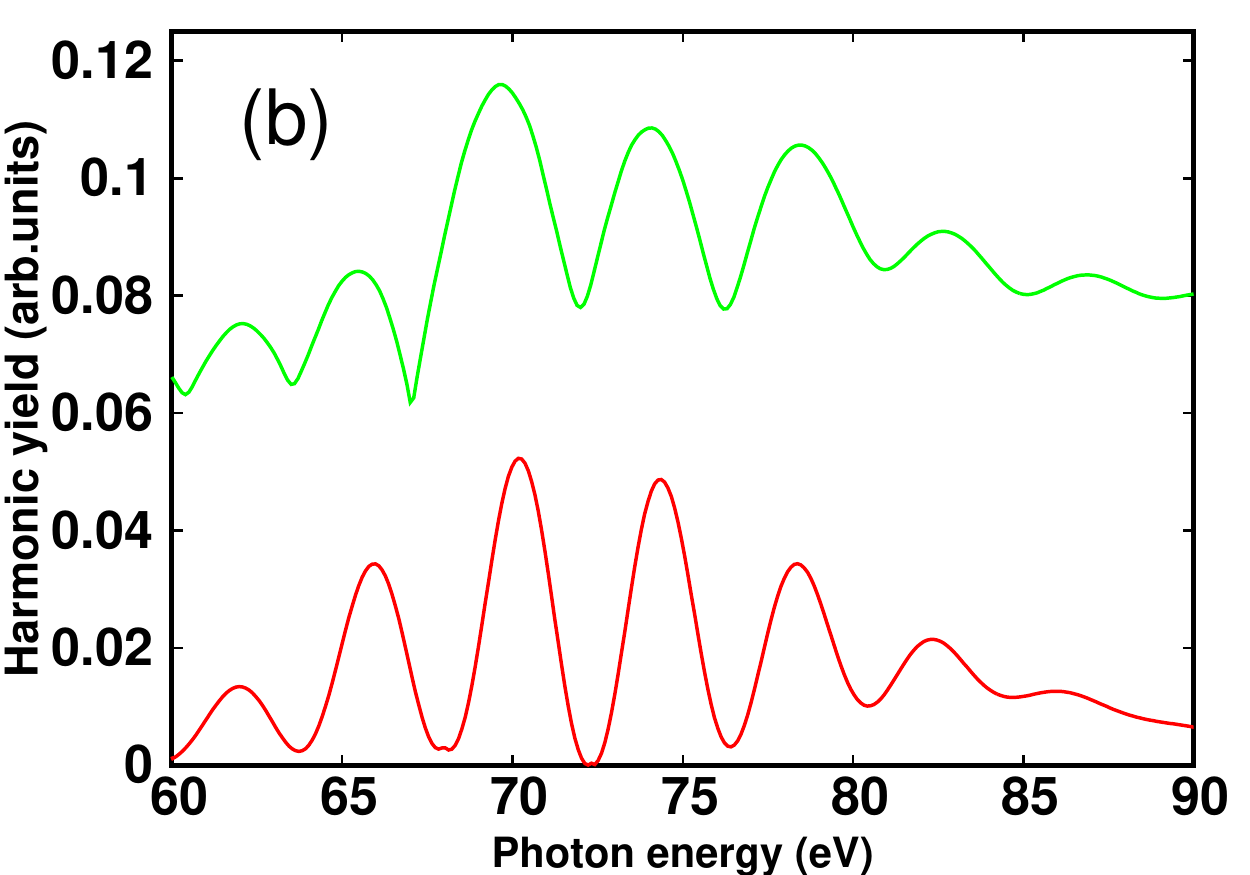}
\caption{\label{fig4} (a)Phase $\phi$ of the harmonics in the XUV region at various number of MLs: 1 ML (orange curve), 10 MLs (green curve), 30 MLs (purple curve) and 50 MLs (black curve). (b) Harmonics calculated from a simple physical model in Eq. (\ref{model}) (see text) (red curve). For reference the HHG spectrum obtained with 30 MLs is also shown with a green curve and with an offset. The parameters of the chirped pulse are: $\lambda_{NIR}=$ 1.27 $\mu m$, $T_c=$ 10 cycles, $\beta=$10.4 rad, $\tau=$ 300 a.u. and $I_{NIR}=$ 2$\times$10$^{13}$ W/cm$^2$.}
\end{figure*}

To provide further insights into the origin of these modulations, we show in Fig. \ref{fig4}(a) the phase information $\phi$ of each harmonic extracted from Eq. (\ref{Dzw}). This  phase is shown for $\beta=$10.4 rad and is displayed at various number of MLs: 1 ML (orange curve), 10 MLs (purple curve), 30 MLs (green curve) and 50 MLs (black curve). Note that the magnitude of the phases is shown with an offset to allow a direct comparison between different results. In Fig. \ref{fig4}(a) an abrupt change of the phase dipole is seen at 30 and 50 MLs, while this behavior is absent in the case of 1 and 10 MLs. This is consistent with the observed spectral modulations in Figs. \ref{fig2}(d) and (e) and their absence in Figs. \ref{fig2}(b) and (c). These observations demonstrate the quantum nature of the modulations, which are further validated by a simple physical model derived from the semi-classical approximation~\cite{Lewenstein1995} given by
\begin{equation}\label{model1}
 H_q(\omega_q) = \int dt e^{-i\omega_q t} \Big[\int_{-\infty}^t dt' d_z(t')F(t') e^{i\phi_q(t,t')} \Big].
\end{equation}
Here the phase $\phi_q(t,t')=-S(t,t')+\Delta E t'$ contains the classic action $S(t)$ and the energy difference $\Delta E$ involving a specific transition. $d_z(t')$ is the transition dipole moment and $F(t')$ is the electric field. The expression in Eq. (\ref{model1}) illustrates that each harmonic $q$ results from a coherent sum of dipole emission amplitudes augmented by the phase $\phi_q$. In a semi-classical picture, this coherent sum involves short and long trajectories followed by the electron during the recombination process~\cite{Lewenstein1995,Bellini1998}. It should be noted that our goal here is not to evaluate the integral in Eq. (\ref{model1}) exactly but rather to provide a simple physical model that captures the basic physics involved in the exact TDSE calculations. At this end, we simplify the expression above by assuming that at each time $t$ of the pulse, the integral over $t'$ is dominated by two trajectories represented by the amplitude $a_i(t)$ and the phase $\phi_i(t)$ ($i=s,l$), where the labels $s$ and $l$ refer, respectively, to the short and long trajectories. Using this assumption leads to  
\begin{equation}\label{model2}
 H_q(\omega_q) = \int dt e^{-i\omega_q t} \Big[ a_s(t) e^{i\phi_s(t)} + a_l(t) e^{i\phi_l(t)}  \Big],
\end{equation}
which can be further simplified as
\begin{equation}\label{model}
 H_q(\omega_q) =  a_s(\omega_q)e^{-i(\Delta\phi_q-\omega_q t_s)} + a_l(\omega_q)e^{i\omega_q t_l}.
\end{equation}
Here $\Delta\phi_q$ is the relative phase between the amplitudes $a_s$ and $a_l$. The times $t_s$ and $t_l$ refer to the short and long travel-time of the electron in the continuum. To evaluate the quantity in Eq. (\ref{model}), we assume that the amplitudes $a_i$ have a Gaussian form $\exp[-\frac{(\omega_q - \omega_i)^2}{2\sigma_i^2}]$ centred around the photon energy of $\omega_i$ and characterized by the width $\sigma_i$. The parameters $\Delta\phi_q$, $\sigma_i$ and $t_i$ are obtained by fitting the model in Eq. (\ref{model}) to the TDSE data. The result stemming from this model is shown in Fig. \ref{fig4}(b) with a red curve and is found to reproduce very well the oscillations imprinted in the spectrum obtained from the exact TDSE calculations. The close agreement between the analytical model in Eq. (\ref{model}) and the TDSE calculations demonstrates the origin of the spectral modulations as a result of quantum interference between short and long trajectories followed by the electron during the recombination and which leads to the generation photons with the same final energy. The emerged quantum interference effects manifest by spectral modulations in HHG spectra, and are of particular interest since they appear in the XUV energy range. This is an interesting finding, which can be exploited for generating isolated attosecond XUV pulses with enhanced intensities.
%------------Begin FIGURE 4
%\begin{SCfigure}[ht]
%\end{SCfigure}
%------------End FIGURE 4
On the other hand, the sensitivity of the generated photons to the number of bulk states is a signature of a coherent control of HHG process. This sensitivity can be understood in the following: during the recombination process the electron wavepacket, which is initially localized close to the surface, recollides not only with surface states but also with different bulk states. These bulk states possess a delocalized character, which makes them a source of additional recollision events, and thus contributing to the efficiency of the HHG process. Therefore the observed pattern in HHG spectra encodes information about the electronic structure of the metal surface, and that makes high-order harmonic spectroscopy induced by means of chirped pulses a powerful characterization tool of materials. Our work therefore introduces a new scheme for coherent control of solid-state HHG that combines the use of optimal chirped pulses with the ability to tune the electronic band of the material itself. 

\section{CONCLUSIONS}\label{conclusions}

In conclusion, we have demonstrated the efficiency of a coherent control scheme applied to enhance HHG from the transition-metal surface Cu(111). This was shown to be achieved on the basis of numerical simulations of the time-dependent Schrodinger equation using a minimal model. Besides showing the extension of the harmonic cutoff by almost a factor of 36, we have demonstrated a control scheme in which the chirped pulse acts as a switch on/off for accessing bulk states, which when getting selectively manipulated by tuning the intrinsic properties of the metal surface itself, it gives rise to quantum modulations in the XUV photon energy region. These emerged quantum effects were linked to the interference between short and long trajectories followed by electrons during the recombination process and which leads to the generation of photons with the same final energy. This interpretation was validated using a simple physical model that contains the basic ingredients necessary for understanding the origin of the quantum nature of the observed effects, and thus providing new insights into the solid-state HHG. Furthermore, it was found that these quantum modulations exhibit a strong sensitivity to the increase of the number of bulk states, and that is a signature that the HHG process encodes information about the intrinsic properties of the metal surface. Our finding thus suggests the use of chirped pulses for spectral characterization of bulk states via high-order harmonic spectroscopy. Most importantly, the results indicate the relevance of metal surfaces when combined with the use of chirped pulses for advancing attosecond science, and potentially for developing compact solid-state based-HHG devices using metal surfaces.

\section*{DATA AVAILABILITY}
The data that support the findings of this study are available from the corresponding author upon reasonable request.

%\bibliography{bibliography.bib}
\providecommand{\noopsort}[1]{}\providecommand{\singleletter}[1]{#1}%

\end{document}